\documentclass[conference]{IEEEtran}
\IEEEoverridecommandlockouts
\usepackage{cite}
\usepackage{amsmath,amssymb,amsfonts, mathtools, stmaryrd}
\usepackage{algorithmic}
\usepackage{graphicx}
\usepackage{textcomp}
\usepackage[nolist]{acronym}
\usepackage{url}

\def\y{{\mathbf y}}

\def\w{{\mathbf w}}

\usepackage[dvipsnames]{xcolor}
\newcommand{\change}[1]{\textcolor{black}{#1}}
\def\ninept{\def\baselinestretch{.95}\let\normalsize\small\normalsize}
\begin{document}

\title{Attention-based distributed speech enhancement for unconstrained microphone arrays with varying number of nodes
\thanks{This work was made with the support of the French National Research Agency, in the framework of the  project DiSCogs “Distant speech communication with heterogeneous unconstrained microphone arrays” (ANR-17-CE23-0026-01). Experiments presented in this paper were partially carried out using the Grid5000 testbed, supported by a scientific interest group hosted by Inria and including CNRS, RENATER and several Universities as well as other organizations (see https://www.grid5000).}
}

\author{\IEEEauthorblockN{Nicolas Furnon}
\IEEEauthorblockA{\textit{Universit{\'{e}} de Lorraine, CNRS, Inria, Loria}\\
F-54000 Nancy, France \\
nicolas.furnon@loria.fr}
\and
\IEEEauthorblockN{Romain Serizel}
\IEEEauthorblockA{\textit{Universit{\'{e}} de Lorraine, CNRS, Inria, Loria}\\
	F-54000 Nancy, France}
\and
\IEEEauthorblockN{Slim Essid}
\IEEEauthorblockA{\textit{LTCI, T\'el\'ecom Paris, Institut Polytechnique de Paris} \\
Palaiseau, France}
\and
\IEEEauthorblockN{Irina Illina}
\IEEEauthorblockA{\textit{Universit{\'{e}} de Lorraine, CNRS, Inria, Loria}\\
	F-54000 Nancy, France}
}
\begin{acronym}
\acro{ds}[DSB]{delay-and-sum beamformer}
\acro{mpdr}[MPDR]{minimum power distortionless response beamformer}
\acro{mvdr}[MVDR]{minimum variance distortionless response beamformer}
\acro{lcmp}[LCMP]{linearly constrained minimum power beamformer}
\acro{lcmv}[LCMV]{linearly constrained minimum variance beamformer}
\acro{mwf}[MWF]{multichannel Wiener filter}
\acro{sdw}[SDW-MWF]{speech distortion weighted multichannel Wiener filter}
\acro{mvdr}[MVDR]{minimum variance distortionless response}
\acro{gevd}[GEVD]{generalized eigenvalue decomposition}
\acro{nmf}[NMF-MWF]{non-negative matrix factorization}
\acro{stft}[STFT]{short-time Fourier transform}
\acroplural{stft}[STFTs]{short-time Fourier transforms}
\acro{tf}[TF]{time-frequency}
\acro{vad}[VAD]{voice activity detector}
\acroplural{vad}[VADs]{voice activity detectors}
\acro{danse}[DANSE]{distributed adaptive node-specific signal estimation}
\acro{mse}[MSE]{mean squared error}
\acro{wasn}[WASN]{wireless acoustic sensor network}
\acroplural{wasn}[WASNs]{wireless acoustic sensor networks}
\acro{doa}[DOA]{direction of arrival}
\acroplural{doa}[DOAs]{directions of arrival}
\acro{irm}[IRM]{ideal ratio mask}
\acroplural{irm}[IRMs]{ideal ratio masks}
\acro{ibm}[IBM]{ideal binary mask}
\acro{dnn}[DNN]{deep neural network}
\acroplural{dnn}[DNNs]{deep neural networks}
\acro{nn}[NN]{neural network}
\acroplural{nn}[NNs]{neural networks}
\acro{lstm}[LSTM]{long short-term memory}
\acro{cnn}[CDNN]{convolutional neural network}
\acroplural{cnn}[CNNs]{convolutional neural networks}
\acro{gru}[GRU]{gated recurrent unit}
\acro{crnn}[CRNN]{convolutional recurrent neural network}
\acro{rnn}[RNN]{recurrent neural network}
\acroplural{rnn}[RNNs]{recurrent neural networks}
\acro{rir}[RIR]{room impulse response}
\acroplural{rir}[RIRs]{room impulse responses}
\acro{ssn}[SSN]{speech shaped noise}
\acro{snr}[SNR]{signal to noise ratio}
\acroplural{snr}[SNRs]{signal to noise ratios}
\acro{sar}[SAR]{source to artifacts ratio}
\acro{sir}[SIR]{source to interferences ratio}
\acroplural{sir}[SIRs]{source to interferences ratios}
\acro{sdr}[SDR]{source to distortion ratio}
\acro{sisdr}[SI-SDR]{scale-invariant signal to distortion ratio}
\acro{stoi}[STOI]{short-time objective intelligibility}

\end{acronym}

\maketitle
\begin{abstract}
	Speech enhancement promises higher efficiency in ad-hoc microphone arrays than in constrained microphone arrays thanks to the wide spatial coverage of the devices in the acoustic scene. However, speech enhancement in ad-hoc microphone arrays still raises many challenges. In particular, the algorithms should be able to handle a variable number of microphones, as some devices in the array might appear or disappear. In this paper, we propose a solution that can efficiently process the spatial information captured by the
	different devices of the microphone array, while being robust to a link failure. To do this, we use an attention mechanism in order to put more weight on the relevant signals sent throughout the array and to neglect the redundant or empty channels.
\end{abstract}
\section{Introduction}
\label{sec:intro}
Ad-hoc microphone arrays are made of several devices like telephones, tablets or hearing aids, each embedded with one or more microphones. They are usually randomly spread in a room, which brings a wide spatial coverage of the room and thus rich recordings of the acoustic scene. Speech enhancement in ad-hoc microphone arrays can benefit from the increased number of microphones in the array and its wide spatial coverage, but it also raises many challenges. In particular, the limited power and computing capacities of the devices, as well as the unconstrained architecture of the array, make it impossible to rely on a fusion center and impose a distributed processing. Besides, the person carrying one of the devices of the array may come in or leave the area covered by the microphone array. This brings the necessity of a flexible processing that can handle a varying number of microphones. Solutions based on a classical signal processing approach have been proposed to alleviate the bandwidth or power constraints \cite{Bertrand2010a, Heusdens2012, OConnor2014}, and some solutions can be used in arbitrary array topologies \cite{Szurley2016, Koutrouvelis2018, Guo2020}. In recent years, solutions based on \acp{dnn} have outperformed the signal-based solutions \cite{Erdogan2016, Heymann2016, Tawara2019}. However, one drawback of \acp{dnn} is that they often require a fixed input dimension, constant at training and testing time, which makes these solutions unflexible to a varying number of channels. A few architectures have been proposed to address the problem of a varying number of microphones. Casebeer et al. for example use recurrent units over the channel axis \cite{Casebeer2018}, but this implies that the order of the input channels is relevant, which can't be guaranteed in real scenarios. Other solutions propose to use shared parameters across input channels and to further fuse the different channels \cite{Luo2020, Wang2020}. These solutions suffer from the drawback that all channels are considered identically by the neural network, whereas they might contain very different information, especially in the context of ad-hoc microphone arrays where the microphones can be wide apart. 

In this paper, \change{we address the typical use case of a person remotely communicating with someone else in a noisy room and recorded by several devices like a telephone or a laptop. Since only the audio signals (and no video) are exchanged between the several devices, we assume that the communication bandwidth is not a limit and that the signals can be exchanged without rate distortion. Rate-constrained speech enhancement in \acp{wasn} remains an issue, especially with low resource devices like hearing aids \cite{Roy2008, Amini2020}. We also assume that the signals are perfectly aligned, although synchronization in \acp{wasn} is an open problem \cite{Schmalenstroeer2015, Chinaev2021}. Our main point of concern is to enhance the speech in a manner that does not rely on a fusion center and remains efficient if some of the recording devices disappear, e.g. if one of them shuts down.
To do so,} we propose a speech enhancement solution that combines classic signal processing with \acp{dnn}. It processes the information captured over the whole microphone array but limits the number of signals exchanged between nodes and operates in arrays with a varying number of devices. This solution is based on our previous work \cite{Furnon2020b}, which benefits from a distributed \ac{mwf} \cite{Bertrand2010a} to alleviate the constraints on the fusion center. \change{It also benefits from the modelling power of \acp{dnn} which proved to efficiently use spatial information for a more precise \ac{tf} mask estimation}. We extend our previous work by designing the \ac{dnn} so that the mask estimation remains accurate while resilient to a link failure. Missing channels are replaced by a constant value indicating a link failure, and an attention mechanism attributes more weight to the relevant channels. We also design an empirical study to clarify the performance improvement brought by the attention mechanism.

This paper is organised as follows. In Section~\ref{sec:problem_formulation}, the problem is formalised. We introduce our solution in Section~\ref{sec:solution} and the experimental setup in Section~\ref{sec:exp_setup}. The results are reported in Section~\ref{sec:results} and Section~\ref{sec:conclusion} concludes this paper.

\section{Problem formulation}\label{sec:problem_formulation}
\subsection{Notations}
We consider $K$ devices, thereafter called nodes. Each node $k$ contains $M_k$ microphones, so that the total number of microphones is $M = \sum_{k=1}^{K}M_k$. 
Following an additive noise model in the \ac{stft} domain, the signal recorded by the $m$-th microphone of the $k$-th node is
$
y_{k, m}(f, t) = s_{k, m}(f, t) + n_{k, m}(f, t)
$
where $s_{k, m}(f, t)$ and $n_{k, m}(f, t)$ are respectively the target speech and the noise recorded by the $m$-th microphone of the $k$-th node at time step $t$ and frequency index $f$. For the sake of conciseness, we will thereafter drop the time and frequency indexes.
The signals recorded by node $k$ are stacked in a vector 
$\y_k~=~[y_{k, 1}, ..., y_{k, M_k}]^T$.
In the following, bold lowercase letters represent vectors. Bold uppercase letters represent matrices. Regular lowercase represent scalars.
The signals recorded by the whole microphone array are stacked into the vector
$
\y~=~[\y_{1}^T, ..., \y_{K}^T]^T
$.

\subsection{Distributed multichannel Wiener filter}\label{subsec:danse}
Bertrand and Moonen introduced the \ac{danse} algorithm which estimates the target speech as recorded by a reference microphone at each node \cite{Bertrand2010a}. At each node, a \ac{mwf} minimizes the mean squared error between the filtered signal and the target speech:
\begin{equation}\label{eq:cost_danse}
	\w_{k} = \mathrm{arg}\min_{\w} \mathbb{E}\{|s_{k, m} - \w^H\tilde{\y}_k|^2\}\,,
\end{equation}
$\mathbb{E}\{\cdot\}$ denotes the expectation and $\cdot^H$ is the Hermitian transpose operator. $\tilde{\y}_k = \left[ \mathbf{y}_k^T,~\mathbf{z}_{-k}^T\right]^T$ gathers the signals of the microphones of node $k$ and the so-called compressed signals $\mathbf{z}_{-k}$ sent by the other nodes: $\mathbf{z}_{-k}~=~[z_1, ..., z_{k-1}, z_{k+1}, ..., z_K]^T$. The compressed signal $z_j$ sent by node $j\ne k$ is the output of a local \ac{mwf} applied at node~$j$:
$
z_{j} =  \mathbf{w}_{jj}^H\mathbf{y}_j
$,
where $\mathbf{w}_{jj} = \mathrm{arg}\min_{\w} \mathbb{E}\{|s_{j, m} - \w^H\y_j|^2\}$.
The solution to Equation \eqref{eq:cost_danse} is given by:
\begin{equation}\label{eq:sol_danse}
	\mathbf{\w_k} = \mathbf{R}_{\tilde{y}\tilde{y}}^{-1}\mathbf{R}_{\tilde{y}s}\mathbf{e}_{k, m}\,,
\end{equation}
where the covariance matrices $\mathbf{R}_{\tilde{y}\tilde{y}}$ and $\mathbf{R}_{\tilde{y}s}$ are estimated from the signals $\tilde{\y}_k$ and $s_{k, m}$ thanks to a \ac{vad} or a \ac{tf} mask, and where $\mathbf{e}_{k, m}$ is a vector of $\mbox{M-1}$ zeros and a $1$ for to the $m$-th microphone of the $k$-th node.
%

This algorithm converges to the centralized node-specific \ac{mwf}, while sparing bandwidth cost as each node sends only one compressed signal to the other nodes \cite{Bertrand2010a}.

\begin{figure}
	\centering
	\includegraphics[width=.8\linewidth, trim=2cm 0 0 0, clip]{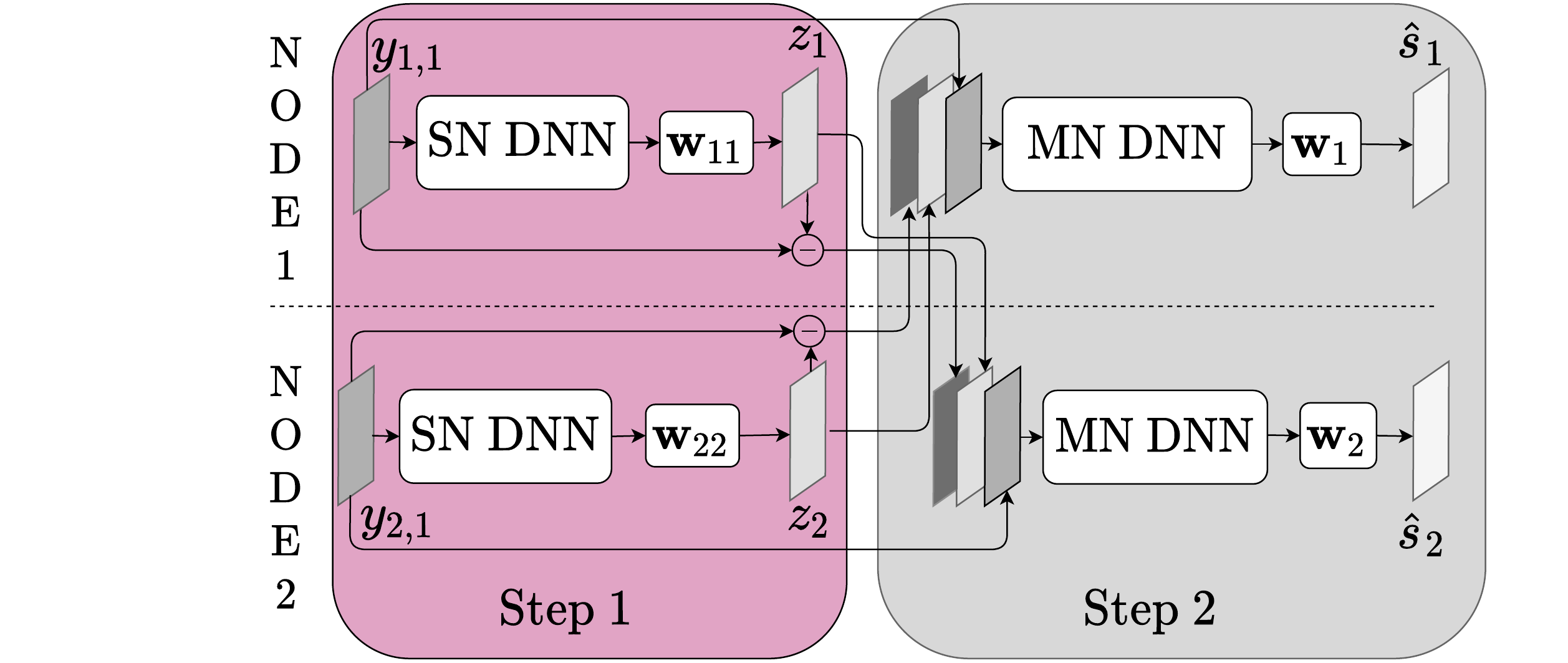}
	\caption{Example of the DNN-based distributed speech enhancement for two nodes, where both the target and the noise estimates are sent as compressed signals. The single-node DNNs (SN-DNN) have access to the local reference only to predict the mask. The multi-node DNNs (MN-DNN) have access to the local reference and the compressed signals to estimate the mask.}
	\label{fig:tango}
\end{figure}

In their paper, Bertrand and Moonen estimate the target speech at node $k$ in an adaptive way where the covariance matrices needed to compute the filter $\w_k$ in Equation~\eqref{eq:sol_danse} are estimated by averaging over time the instantaneous spatial covariance matrices. This however raises stability issues that are beyond the scope of this paper. Thus, we split the adaptive process of \ac{danse} into two distinct steps represented in Figure~\ref{fig:tango}. The first step computes the compressed signals and the second step estimates the target speech signal once every node has received the compressed signals of the other nodes. 

\subsection{DNN-based distributed multichannel Wiener filter}
In previous work \cite{Furnon2020}, we replaced the oracle \ac{vad} used in \ac{danse} by a \ac{tf} mask predicted by a \ac{crnn}\change{, in a similar manner as \cite{Erdogan2016, Heymann2016}}. We showed that the compressed signals sent to compute the filter of Equation~\eqref{eq:sol_danse} could also help to improve the mask prediction at the second step by a multi-node \ac{dnn}. This achieved better performance than with an oracle \ac{vad}. In an extended study, we generalized these results to real-life scenarios and showed that sending the noise estimate rather than the target estimate could improve the performance depending on the \ac{sir} at the receiving node \cite{Furnon2020b}. To take full advantage from the spatial coverage of the distributed microphone array, in the following, both the target and noise\footnote{Assuming a noise additive model, the compressed noise $\tilde{n}_k$ at node $k$ is estimated as $\tilde{n}_k = y_{k, m} - z_k$.} estimates will be sent as represented in Figure~\ref{fig:tango}. 

\section{Attention-based distributed speech enhancement algorithm}\label{sec:solution}
\begin{figure}
	\centering
	\includegraphics[width=.75\linewidth]{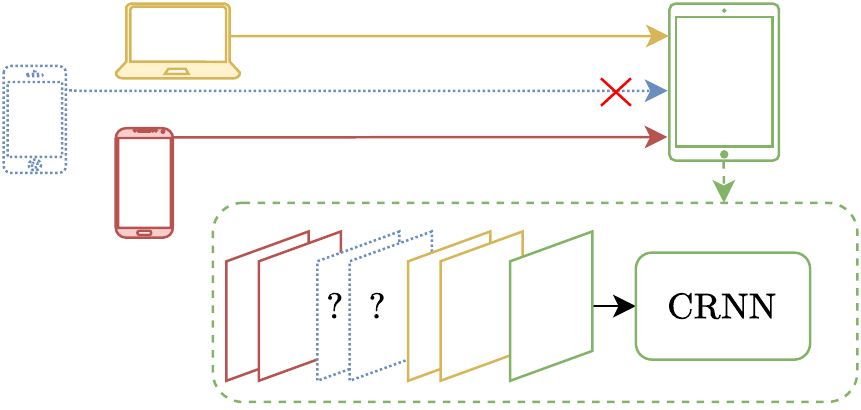}
	\caption{Schematization of a situation where three nodes are expected to send compressed signals, but only two nodes actually send them. The \ac{crnn} expects the local (green) mixture and the target and noise estimates of all distant nodes (yellow, blue, red).}
	\label{fig:broken_link}
\end{figure}

In this paper, we focus on the typical problem of a node disappearing from the microphone array, for example because the owner of the corresponding device leaves the room. Such a case is schematized in Figure~\ref{fig:broken_link}. The model architecture used in our previous work requires a constant number of input channels so it cannot be used for a variable number of nodes. In the context of broken or disappearing links, this raises an issue which we address in this paper. The contribution of this paper is twofold. First, we propose a solution to deal with a variable number of nodes which is robust to broken links. Second, we examine the performance of this solution with an empirical study in order to explain the obtained performance.

\begin{figure}
	\centering
	\includegraphics[width=.8\linewidth, trim=0 1.5cm 1cm 0, clip]{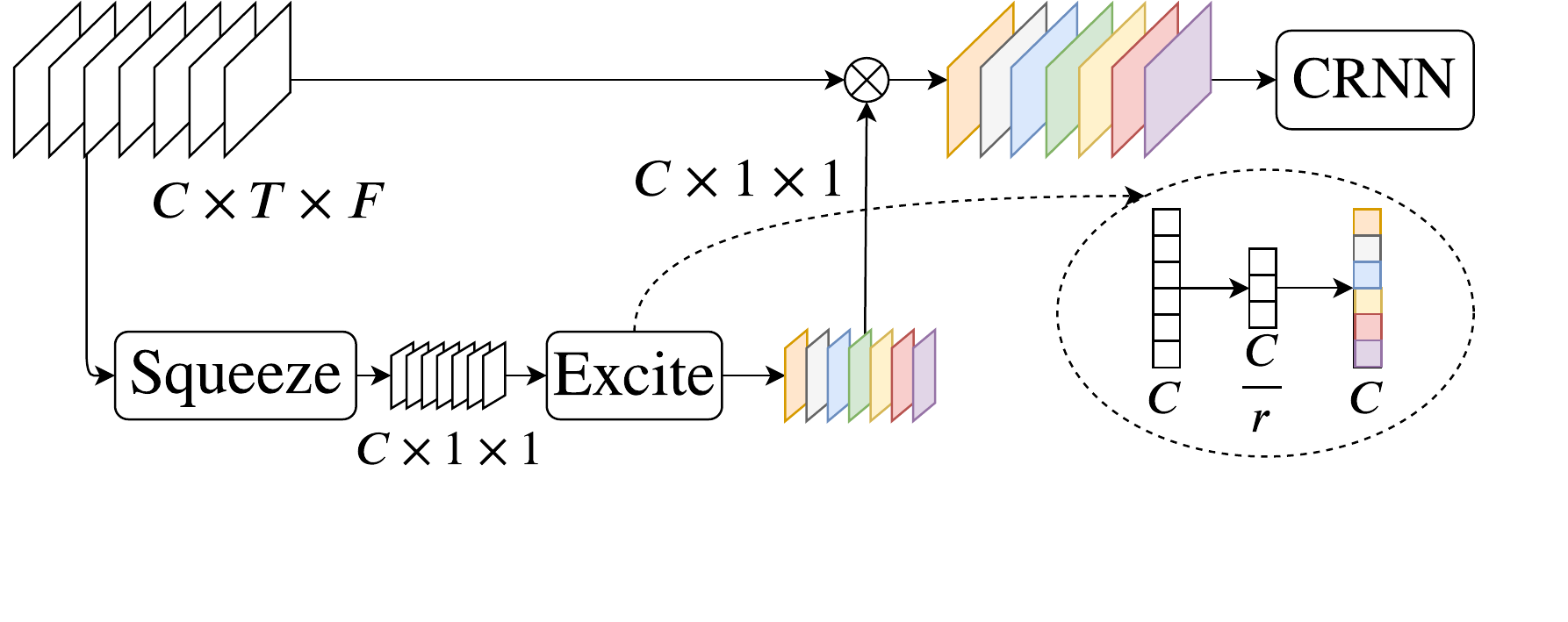}
	\caption{Illustration of the \textit{Squeeze-and-Excitation} attention mechanism. $C$, $T$ and $F$ denote the number of channels, time frames and frequency bins respectively. $r$ is the reduction ratio.}
	\label{fig:se_crnn}
\end{figure}

To cope with a variable number of nodes, we fix the number of input channels to a constant maximal number. If a node disappears, we replace the corresponding unreceived signals with a small constant negative value, which symbolises a broken link. \change{While this is limited by the maximal number of devices considered, it seems a valid solution in scenarios where a small number of devices already captures most of the spatial information.} We propose to use an attention mechanism to force the \ac{dnn} to consider differently the input channels. This is illustrated in Figure \ref{fig:se_crnn}. The attention mechanism is a \textit{Squeeze-and-Excitation} (SE) block introduced by Hu et al.~\cite{Hu2018}. The mechanism operates in two steps. In the first step, it \textit{squeezes} the input tensor over the time and frequency axis to output a one-dimensional vector. The squeezing operation is an average pooling that enables to compress the whole spatial information into one bin. It embeds the input data into a global vector so that contextual information can be exploited in the second step. In the second step, the one-dimensional vector is passed to a multilayer perceptron composed of two fully-connected layers. These layers form a bottleneck where the input dimension is reduced by a factor $r$ in order to reduce the complexity of the mechanism and to prevent overfitting. This mechanism was shown to exploit contextual information and dependencies over the channel axis while limiting the complexity of the model~\cite{Hu2018}. In the sequel, we will refer to the output of the SE mechanism as weights.

\section{Experimental setup}\label{sec:exp_setup}
\subsection{Models}\label{subsec:models}
Four models are compared in order to study the performance of our proposed approach. 
The first model is a single-node \ac{crnn} whose structure is described in Section \ref{subsec:setup}. It has only access to the local reference signal to estimate the mask at the second step of the algorithm. This is the simplest method that is invariant to the number of nodes, since it does not rely on the compressed signals to predict the mask. It is denoted ``SN''. The second model is the model that has been used in our previous experiments \cite{Furnon2020b}. It is a multi-node neural network that estimates the mask based on the local reference signal and the compressed signals sent from the other nodes. At inference time, its missing channels are replaced by a constant negative value, but during the training, all the links were always valid. It is denoted by ``MN$_{0}$''. The third model is the same architecture as the second model, but each training sample could contain 0 to 3 broken links. It is denoted by ``MN$_{0-3}$''. The fourth model is our proposed solution with a SE mechanism at its input. It is trained with 0 to 3 broken links at every training sample. It is denoted by ``MN-SE''.

\subsection{Dataset}\label{subsec:dataset}
The dataset used to train and test our proposed solution is the same as the one of our previous work \cite{Furnon2020b}. It consists of simulations of typical shoebox-like rooms with one target source and one noise source randomly laid in the room. Four nodes of four microphones each are also randomly placed in the room. All sources and nodes are distant of at least 50~cm of the closest source, node and wall.

The speech material is taken from LibriSpeech~\cite{Panayotov2015}. The noise material is downloaded from Freesound~\cite{Freesound}. It is split into two non-overlapping subsets of Freesound users for the training and testing sets. Some speech-shaped noise was also used to train the \ac{dnn} because it was shown to improve the robustness of the \ac{dnn} \cite{Furnon2020b}.

The rooms were simulated with the Python toolbox Pyroomacoustics \cite{Scheibler2018}. The \ac{sir} of the non-reverberated source signals is randomly taken between 0~dB and 6~dB. \change{The reverberation time ranges from 150~ms to 400~ms.} We created around 25~hours of training material and 2.5~hours of testing material\footnote{The code to generate the dataset is publicly available at \url{https://github.com/nfurnon/disco/tree/master/dataset_generation}.}.

\subsection{Setup}\label{subsec:setup}
All the signals are sampled at 16~kHz. The \ac{stft} is computed with a Hann window of 32~ms with an overlap of 16~ms. The \ac{crnn} architecture is composed of three convolutional layers followed by a recurrent layer and a fully-connected layer. The convolutional layers have 32, 64 and 64 filters, with kernel size $3 \times 3$ and stride $1 \times 1$. Each convolutional layer is followed by a batch normalisation and a maximum-pooling layer of kernel size $4 \times 1$ so that no pooling is applied over the time axis. The recurrent layer is a 256-unit GRU. The fully-connected layer has 257 units with a sigmoid activation function. The reduction ratio in the excitation operation is set to 2. The input of the model are the magnitudes of the \ac{stft} windows of 21 consecutive frames and the ground truth labels are the corresponding frames of the ideal ratio mask. \change{At test time, only the middle frame of the predicted window is considered to estimate the mask, so sliding windows of the input are fed to the \ac{dnn}. The mask of the whole signal is predicted before being used to enhance the speech in a batch mode.} When a link is broken between a node and the rest of the array, both the target and the noise magnitudes which are missing are replaced by an array equal to $-10^{-7}$ in all \ac{tf} bins. This way, the \ac{dnn} always has 7 input channels (one local signal plus $3\times 2$ compressed signals) whatever the number of nodes it is connected to.

To better analyse the impact of missing channels on the \ac{dnn} performance, we only consider missing channels at the input of the \ac{dnn} and still use all the compressed signals at the filtering operations. This is purely artificial, since available signals at the filtering operations should also be available to predict the mask. However, this setup allows us to disentangle the impact of a broken link on the \ac{dnn} and on the \ac{mwf}. We can then analyse the performance from a \ac{dnn} point of view which is the focus of this paper\footnote{Preliminary studies showed that missing channels at the filtering operation lead to a limited drop of performance.}. 

\subsection{Performance evaluation}\label{subsec:performance_evaluation}
Three metrics are used to evaluate the results: the \ac{sir} improvement \cite{Vincent2006}, denoted as $\Delta$SIR; the \ac{sar}; and the \ac{stoi} improvement \cite{Taal2010}, denoted as $\Delta$STOI. The references needed to compute these metrics are the non-reverberated noise and speech signals. 
All the reported results correspond to the average metric over all the nodes of all configurations of the test set. This was decided in order to report the overall performance in the microphone array.

\section{Results and analysis}\label{sec:results}
\subsection{Resilience to missing channels}\label{subsec:resilience}
We compare the four models introduced in Section~\ref{subsec:models} on the testing set and report the results in Figure~\ref{fig:results}.
Replacing the missing channels by a fixed value can help the \ac{dnn} to be resilient to link failures, at the condition that this \ac{dnn} was trained to deal with such signals (MN$_{0-3}$, MN-SE). The second \ac{dnn} (MN$_{0}$), which was not trained with broken links, fails to enhance the speech as soon as one of the four nodes is disconnected from the rest of the microphone array. If the \ac{dnn} is trained to deal with missing channels (MN$_{0-3}$), it can still exploit the spatial information sent from the other nodes since MN$_{0-3}$ outperforms the single-node \ac{dnn} when 2 or fewer links are broken. When all nodes are disconnected ($L=3$), the noise reduction is lower than with the single-node \ac{crnn}, because the amount of missing data is too high for the MN$_{0-3}$. Still, the performance in terms of \ac{sar} and \ac{stoi} is equal between the two models. With an attention mechanism (MN-SE), the noise reduction also decreases when the number of broken links increases, but to a lesser extent than for the other models, and it always significantly outperforms, in terms of \ac{sir} and \ac{sar}, both the single-node \ac{dnn} and the multi-node \ac{dnn} trained with missing channels. Besides, the \ac{sar} with MN-SE is constant over $L$, which shows that using this model does not introduce artefacts although channels are missing. This means that the SE mechanism not only helps to exploit the spatial information that is actually received, but also increases the performance of the \ac{crnn} even when no compressed signal is received ($L=3$). Lastly, the increased difference between the performance of MN$_{0-3}$ and MN-SE when $L$ increases indicates a stronger resilience of MN-SE compared to the former model.
\newcommand\len{.8}
\newcommand\hei{.55}
\begin{figure}
	\centering
	\includegraphics[width=\len\linewidth, height=\hei\linewidth, trim=.4cm .4cm .35cm 0, clip]{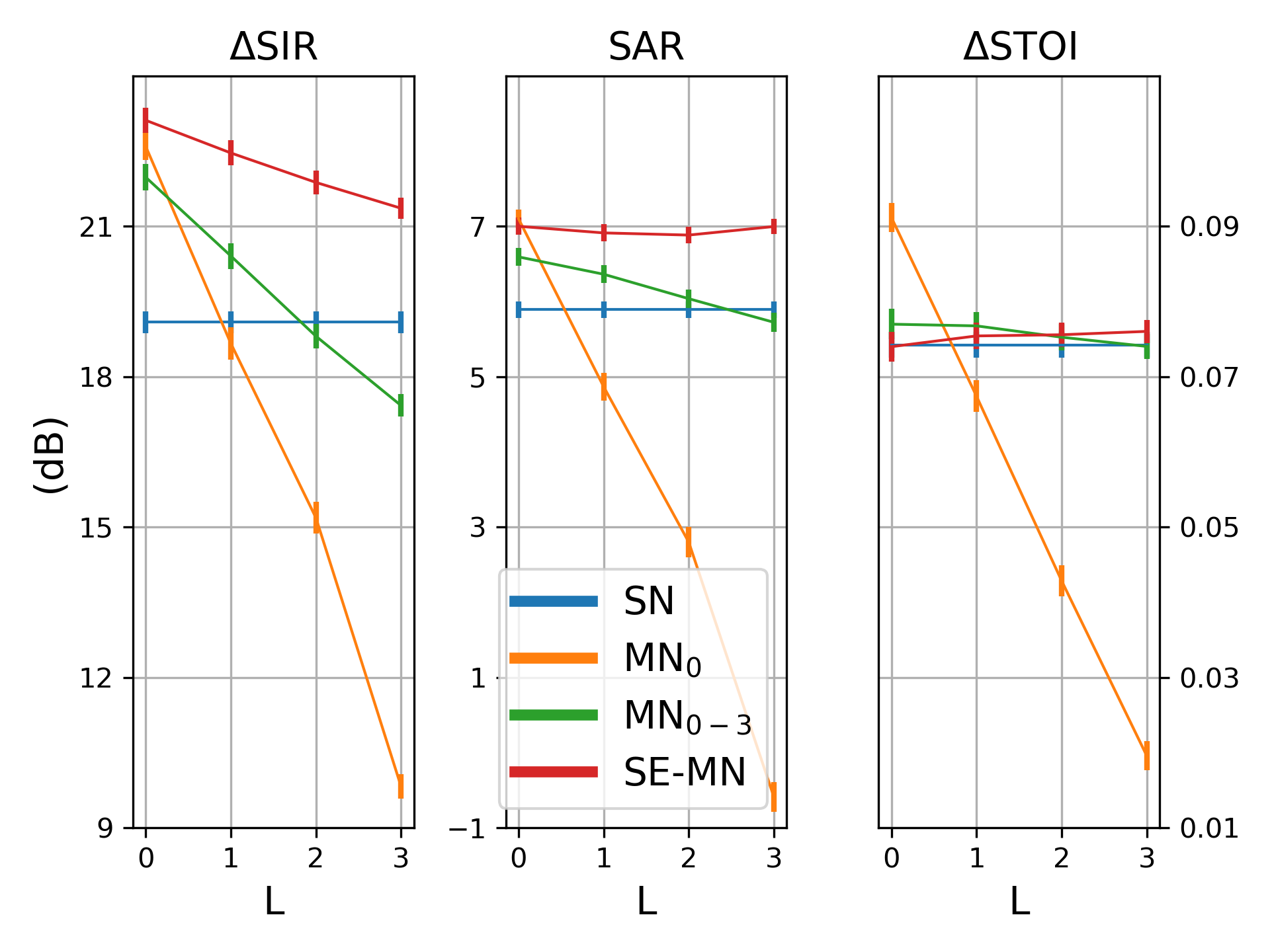}
	\caption{Performance over link failures. $L$ refers to the number of broken links. The error bars represent the 95\% confidence intervals.}
	\label{fig:results}
\end{figure}

\subsection{Dissociation of the effects of the attention mechanism}\label{subsec:ablation}
\begin{figure}
	\centering
	\includegraphics[width=\len\linewidth, height=\hei\linewidth, trim=.4cm .4cm .35cm 0, clip]{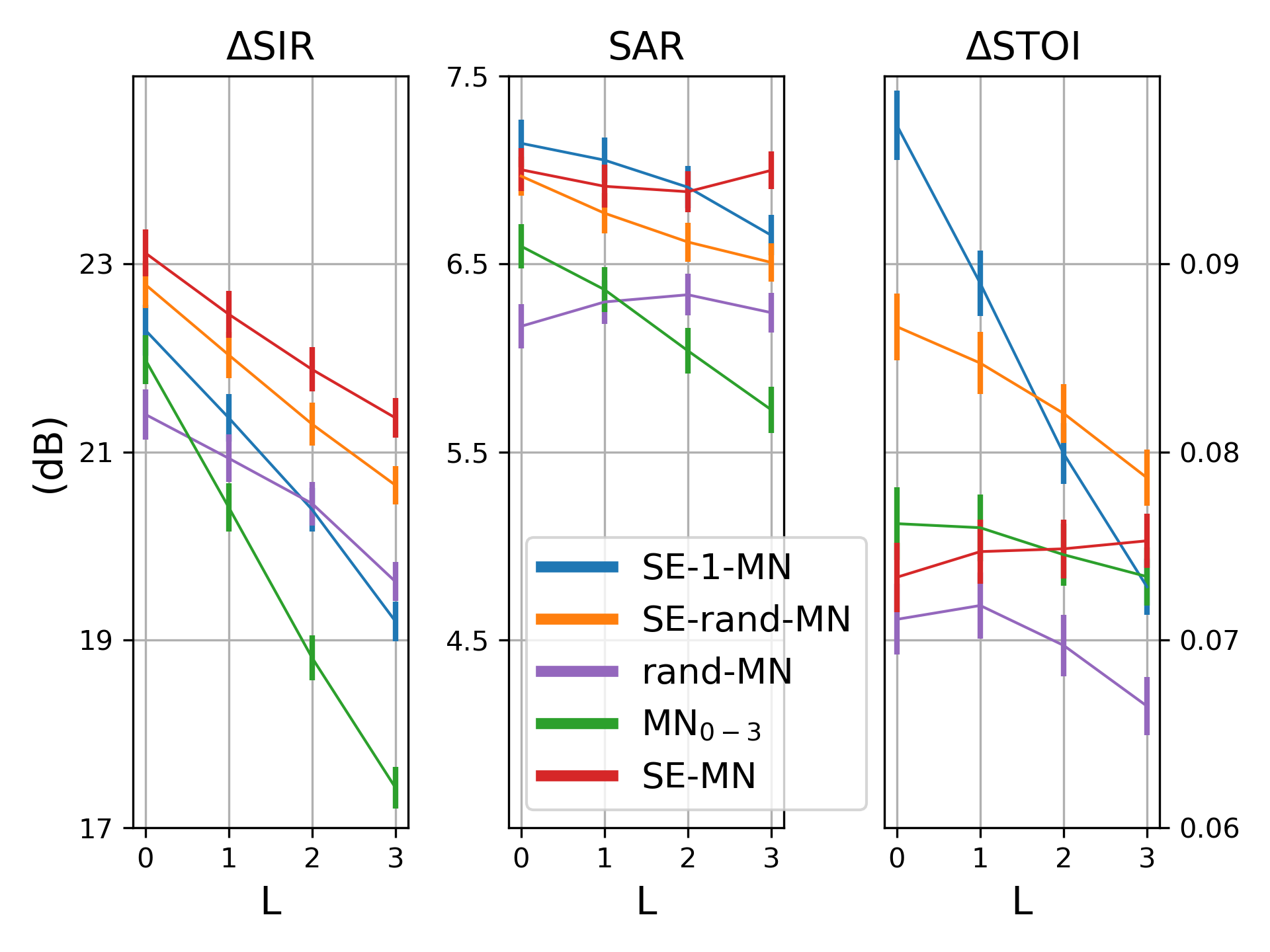}
	\caption{Performance over link failures when the effects of the SE mechanism and of the weights are dissociated. $L$ refers to the number of broken links. The error bars represent the 95\% confidence intervals.}
	\label{fig:ablation}
\end{figure}

In this section, we propose an ablation study to disentangle the effects of the SE branch and the effects of the weights. The performance improvement can have two reasons. The first reason is that the weights applied to the channels highlight the channels of interest. The second reason is that the SE branch helps the whole \ac{dnn} to train better. Considering these two hypotheses, we train the following three \acp{dnn}:
\begin{itemize}
	\item rand-MN; a multi-node \ac{crnn} without SE mechanism, and with random weights applied on the input channels.
	\item SE-rand-MN; a multi-node \ac{crnn} with SE mechanism but whose attention weights are replaced by random values at train time and at inference time.
	\item SE-1-MN; a multi-node \ac{crnn} with SE mechanism but whose attention weights are replaced by 1 at train time and at inference time.
\end{itemize}
The results of these models, together with the results of the previous models MN-SE and MN$_{0-3}$, are represented in Figure~\ref{fig:ablation}.
The impact of the SE module alone can be studied by comparing MN-SE-rand with MN-rand and MN-SE with MN$_{0-3}$. In both cases the SE module helps improving the overall performance and the robustness of the model. The impact of the weights alone is more complex to analyse. Comparing MN-rand with MN$_{0-3}$ and MN-SE-rand with MN-SE-1 helps understanding the influence of applying any weights on the input of the \ac{crnn}. Although this brings worse performance in terms of \ac{stoi}, it increases the robustness of the models. Indeed, when $L$ increases, the performance of the models with weights decreases much less than the performance of the models without weights. Lastly, comparing MN-SE-1 with MN-SE helps analysing the influence of applying the correct weights on the input of the \ac{crnn}. From the results, using the correct weights increases the \ac{sir} and \ac{sar} but lowers the \ac{stoi}. However, the model without weights is much less robust to missing channels, as the \ac{stoi} decreases drastically when $L$ increases. To sum up, the SE branch helps increasing the performance of the whole model, even with random weights applied on the input of the \ac{crnn}. Using the correct values of the weights is important primarily for a higher number of missing channels.

\section{Conclusion}\label{sec:conclusion}
We introduced a distributed multichannel speech enhancement algorithm that handles a varying number of input channels. Based on an attention mechanism, it exploits the spatial information and minimizes the performance drop due to link failures. An ablation study led to the conclusion that the SE mechanism helps to improve the performance of the whole network, and that the weights importance increases with the number of missing channels. Future works foresees to analyse the behaviour of the proposed system when the assumptions about the bit-rate and synchronization do not hold.

\bibliographystyle{IEEEbib}
\bibliography{refs}

\end{document}